\documentclass{aastex631}

\begin{document}

\title{An Extreme Stellar Prominence Eruption Observed by LAMOST Time-Domain Spectroscopy}

\author{Hong-Peng Lu}
\affiliation{State Key Laboratory of Public Big Data, Guizhou Radio Astronomical Observatory, and College of Physics, Guizhou University, Guiyang 550025, People's Republic of China}
\affiliation{School of Earth and Space Sciences, Peking University, Beijing 100871, People's Republic of China}

\author{Hui Tian}
\altaffiliation{huitian@pku.edu.cn}
\affiliation{School of Earth and Space Sciences, Peking University, Beijing 100871, People's Republic of China}

\author{Li-Yun Zhang}
\affiliation{College of Physics, Guizhou University, Guiyang 550025, People's Republic of China}

\author{He-Chao Chen}
\affiliation{School of Physics and Astronomy, Yunnan University, Kunming 650500, People's Republic of China}

\author{Ying Li}
\affiliation{Key Laboratory of Dark Matter and Space Astronomy, Purple Mountain Observatory, Chinese Academy of Sciences, Nanjing 210023, People's Republic of China}

\author{Zi-Hao Yang}
\affiliation{School of Earth and Space Sciences, Peking University, Beijing 100871, People's Republic of China}

\author{Jia-Sheng Wang}
\affiliation{School of Earth and Space Sciences, Peking University, Beijing 100871, People's Republic of China}

\author{Jia-Le Zhang}
\affiliation{School of Earth and Space Sciences, Peking University, Beijing 100871, People's Republic of China}
\affiliation{ASTRON, Netherlands Institute for Radio Astronomy, Oude Hoogeveensedijk 4, Dwingeloo, 7991 PD, The Netherlands}
\affiliation{Kapteyn Astronomical Institute, University of Groningen, P.O. Box 800, 9700 AV, Groningen, The Netherlands}

\author{Zheng Sun}
\affiliation{School of Earth and Space Sciences, Peking University, Beijing 100871, People's Republic of China}



\begin{abstract}

We report the detection of an extreme stellar prominence eruption on the M dwarf LAMOST J044431.62+235627.9, observed through time-domain H$\alpha$ spectroscopy with the Large Sky Area Multi-Object Fiber Spectroscopic Telescope (LAMOST). This prominence eruption was accompanied by a superflare lasting over 160.4 minutes. The H$\alpha$ line profile exhibits significant blue-wing enhancement during the impulsive phase and near the flare peak, with a projected bulk blueshift velocity of $-228\pm11$~km~s$^{-1}$ and a maximum blueshift velocity reaching $-605\pm15$~km~s$^{-1}$. Velocity analysis of the eruptive prominence at various heights above the stellar surface indicates that some of the projected ejection velocities along the line of sight exceed the corresponding escape velocities, suggesting a potential coronal mass ejection (CME). The equivalent width (EW) of the H$\alpha$ blue-wing enhancement in this eruption appears to be the largest observed to date and is comparable to the EW of the H$\alpha$ line profile during the quiescent phase of the host star. We performed a two-cloud modeling for the prominence and the associated flare, which suggests that the eruptive prominence has a mass ranging from $1.6 \times 10^{19}~\text{g}$ to $7.2 \times 10^{19}~\text{g}$. More importantly, the mass ratio of the erupting prominence to its host star is the largest among all reported stellar prominence eruptions/CMEs.

\end{abstract}



\section{Introduction} \label{sec:intro}

Coronal mass ejections (CMEs), involving the ejection of magnetized plasma from the stellar coronae, are among the most energetic eruptive events on active stars. They could contribute to the angular momentum loss of stars, lead to atmospheric escape and composition change, and thus affect the habitability of orbiting exoplanets \citep{2016NatGe...9..452A, 2019LNP...955.....L, 2024ApJ...971..153X}. While CMEs have been extensively studied on the Sun \citep{2000JGR...10523153F, 2004ApJ...602..422L, 2010SunGe...5....7G, 2011LRSP....8....1C}, direct detections of stellar CMEs are sparse due to observational limitations \citep{2019ApJ...877..105M}. 

Currently, stellar CME detection relies primarily on several methods \citep{2022SerAJ.205....1L, 2023ScSnT..53.2021T, 2024Univ...10..313V}, including the coronal dimming method \citep[e.g.,][]{2014ApJ...789...61M, 2021NatAs...5..697V, 2022ApJ...936..170L, 2022ApJ...928..154J, 2024ApJ...970...60X}, the radio bursts method \citep[e.g.,][]{2020ApJ...905...23Z, 2024A&A...686A..51M}, and the Doppler-shift method. The Doppler-shift method directly traces CME plasma motion by detecting wing asymmetries or Doppler shifts in spectral lines. This method has been applied across various wavelengths to detect stellar CMEs. For example, using Balmer lines, a stellar filament eruption on the solar-like G-type star EK Dra was detected via blue-wing absorption in the H$\alpha$ line \citep{2022NatAs...6..241N}. Additionally, CME candidates associated with prominence eruptions have been detected on several late-type main-sequence stars through red or blue wing enhancements in Balmer lines \citep[e.g.,][]{2018A&A...615A..14F, 2019A&A...623A..49V, 2020MNRAS.499.5047M, 2021ApJ...916...92W, 2021PASJ...73...44M, 2022A&A...663A.140L, 2022ApJ...928..180W, 2023ApJ...948....9I, 2024ApJ...963...13C, 2024ApJ...961..189N, 2024MNRAS.532.1486L}. In the X-ray band, CME candidates have been detected through asymmetries in high-temperature coronal lines during stellar flares \citep[e.g.,][]{2019NatAs...3..742A, 2022ApJ...933...92C, 2024ApJ...969L..12I}. Moreover, asymmetries in transition region lines observed in the far ultraviolet (FUV) during stellar flares have also been linked to possible CMEs \citep[e.g.,][]{2011A&A...536A..62L}.

The Sun, being the closest star to Earth, serves as a critical reference for guiding the detection of CMEs on other stars through Sun-as-a-star spectral observations. Sun-as-a-star spectra in the Extreme Ultraviolet (EUV) and H$\alpha$ bands have revealed blue wing asymmetries during solar prominence or filament eruptions \citep[e.g.,][]{2022ApJ...931...76X,  2023ApJ...953...68L, 2024ApJ...964...75O}. Based on Sun-as-a-star analyses, \cite{2022ApJS..260...36Y, 2024ApJ...966...24Y} developed an analytical CME model and demonstrated that stellar CMEs could be detected through the Doppler shifts or profile asymmetries of EUV spectral lines.

Full-phase time-domain spectroscopic observations during stellar flares/CMEs are very rare. And most stellar CME candidates detected using the Doppler-shift method do not exceed the stellar surface escape velocities, making it difficult to confirm them as definitive stellar CMEs. In this letter, we report an extreme prominence eruption on the M dwarf LAMOST J044431.62+235627.9, identified through blue-wing enhancement in the H$\alpha$ line from LAMOST observations. The time-domain spectroscopic observation covers different phases of the accompanied superflare, with the line-of-sight projected velocities of some parts of the erupting prominence exceeding the escape velocity. More importantly, the mass ratio of this erupting prominence to its host star is the largest among all reported stellar prominence eruptions/CMEs. Section 2 describes the observations and data reduction. Section 3 presents an analysis of the asymmetries in H$\alpha$ line profiles. Section 4 covers the statistical analysis of H$\alpha$ blue-wing asymmetry during the stellar flares. Section 5 explores the blue-wing asymmetry using the two-cloud model. And finally, a brief summary is provided.

\section{Observation and Data Reduction} \label{sec:sec2}

LAMOST J044431.62+235627.9 (also known as TIC 385505906) is a highly active M-type main-sequence star. The Transiting Exoplanet Survey Satellite \citep[TESS;][]{2015JATIS...1a4003R} observed this M dwarf twice in 2021 (sectors 43 and 44) and detected three superflares, corresponding to an occurrence rate of one superflare every 14.6 days (see Appendix \ref{sec:1A}). The projected rotational velocity ($v \sin i$) of this M dwarf is $ 7.1 \, \mathrm{km \, s^{-1}}$, and its stellar age is estimated to be 690 Myr \citep{2018A&A...616A..10G, 2021A&A...645A..42I}. LAMOST low-dispersion spectroscopic observations classify it as an M4-type star, with a stellar surface effective temperature of $3283\pm117$~K and a surface gravity of $\log(g/\mathrm{cm\,s^{-2}})$ = $5.19\pm0.27$. According to \cite{2023AJ....165..267H}, the star has an apparent V-band magnitude of 15.61, is located at a distance of $51.04\pm0.06$~pc from Earth, with a stellar luminosity of $0.0109\pm0.0003$~L$_\odot$, a stellar radius of $0.34\pm0.01$~R$_\odot$, and a stellar mass of $0.32\pm0.01$~M$_\odot$.
 
Since October 2018, LAMOST has been conducting the medium-resolution time-domain spectroscopic survey (LAMOST-MRS) \citep{2012RAA....12.1197C, 2012RAA....12.1243L, 2012RAA....12..723Z, 2019ApJS..244...27W, 2020arXiv200507210L}. The raw data from LAMOST-MRS are processed using the LAMOST 2D pipeline, which includes standard procedures such as dark and bias subtraction, flat-field correction, spectral extraction, sky subtraction, and wavelength calibration \citep{2015RAA....15.1095L, 2020ApJS..251...15Z}. The LAMOST-MRS spectra cover a blue arm (4950--5350 \AA) and a red arm (6300--6800 \AA), both with a resolution of 7500 at 5163 \AA{} and 6593 \AA{}, respectively. The typical radial velocity accuracy of LAMOST-MRS observations is 1 km s$^{-1}$ \citep{2019RAA....19...75L}.

\begin{figure}[ht!]
\plotone{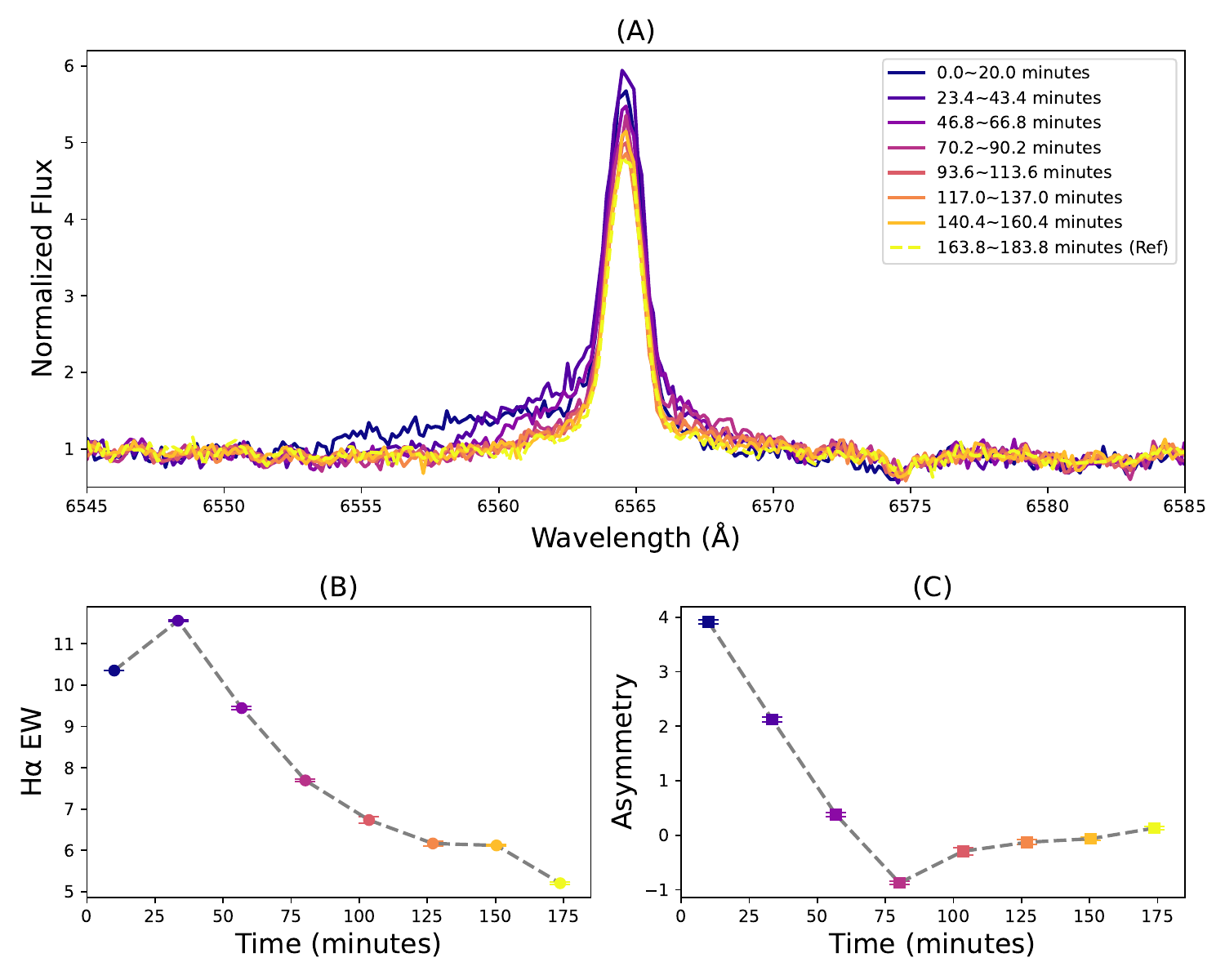}
\caption{Evolution of the H$\alpha$ line profile during a superflare on the M-type dwarf LAMOST J044431.62+235627.9. Panel (A) shows the normalized H$\alpha$ line profiles, with different colors representing spectra observed at different times. Panel (B) presents the time evolution of the H$\alpha$ equivalent width. Panel (C) illustrates the difference between the integrated fluxes of the blue wing (6552.6\,--\,6564.6\,\AA; H$\alpha$\_blue) and red wing (6564.6\,--\,6576.6\,\AA; H$\alpha$\_red) of the normalized H$\alpha$ line profile (H$\alpha$\_blue - H$\alpha$\_red).
\label{fig:haevolution}}
\end{figure}

On December 14, 2018, at 13:54:51.0 UT, LAMOST began a time-domain spectroscopic observation of LAMOST J044431.62+235627.9, obtaining eight spectra with an exposure time of 1200 seconds each. Due to the low signal-to-noise ratio in the blue arm, we only analyzed the red arm spectra. We used the ‘laspec’ toolkit \citep{2021ApJS..256...14Z} for continuum normalization and cosmic ray removal. The processed time-domain H$\alpha$ line profiles are shown in Figure \ref{fig:haevolution}(A). Figure \ref{fig:haevolution}(A) reveals significant intensity variations in the H$\alpha$ line core and, more importantly, noticeable blue-wing enhancements in the H$\alpha$ line profile. Figure \ref{fig:haevolution}(B) shows the time evolution of the equivalent width (EW) of the H$\alpha$ line, calculated over the wavelength range of 6564.6 $\pm$ 12 \AA. The rest wavelength of the H$\alpha$ line in air, as provided by LAMOST, is 6564.6~\AA. Figure \ref{fig:haevolution}(B) indicates that the time-domain spectra capture the impulsive phase, peak, and decay phase of the flare. Additionally, we integrated the red wing (6564.6--6576.6 \AA) and the blue wing (6552.6--6564.6 \AA) of the H$\alpha$ line separately, then calculated the difference (H$\alpha$$_\mathrm{blue}$ - H$\alpha$$_\mathrm{red}$), as shown in Figure \ref{fig:haevolution}(C). Figure \ref{fig:haevolution}(C) demonstrates that the first three H$\alpha$ line profiles exhibit blue-wing enhancements, which gradually decrease, followed by red-wing enhancements in the subsequent three profiles, which also gradually diminish. The final two H$\alpha$ line profiles show no significant asymmetry, with the last profile having the lowest EW. Therefore, we selected the last H$\alpha$ spectrum as the reference spectrum, representing the quiescent state of the host star. The first seven spectra correspond to the stellar flare, which lasted 160.4 minutes. By subtracting the reference spectrum from the flare spectra, we obtained the change in H$\alpha$ EW induced by the flare. Using this, in conjunction with the stellar bolometric luminosity ($L_{bol}$) and the ratio ($\chi_\mathrm{Ha}$) between the continuum flux near the H$\alpha$ line and the bolometric flux \citep{2004AJ....128..426W, 2018MNRAS.476..908F}, we estimated the flare energy in the H$\alpha$ band to exceed $4.6 \times 10^{31}$~erg using the following equation:
\begin{equation}
E_{H\alpha } = \chi _{H\alpha } \cdot L_{bol} \cdot \int EW_{H\alpha }(t)\, dt
\end{equation}
Studies of superflares on G-type main sequence stars \citep{2015EP&S...67...59M} and flares on M dwarfs \citep{2014ApJ...797..121H} indicates that the flare duration is a function of flare energy ($\tau_{flare} \propto E_{flare}^{\alpha}$), and the $\alpha$ value for both types of stellar flares are similar, suggesting that all these flares are likely caused by the same basic physical process of magnetic reconnection \citep{2016ApJ...829..129S}. Based on the results of \cite{2015EP&S...67...59M}, the e-folding decay time of our flare is 50.3 minutes, and the corresponding flare bolometric energy is estimated to be $3.53 \times 10^{35}$~erg, classifying it as a superflare.

\section{Asymmetry Analysis of H\texorpdfstring{$\alpha$}{alpha} Line Profiles}
 \label{sec:sec3}

\begin{figure}[ht!]
\centering
\includegraphics[width=0.7\textwidth]{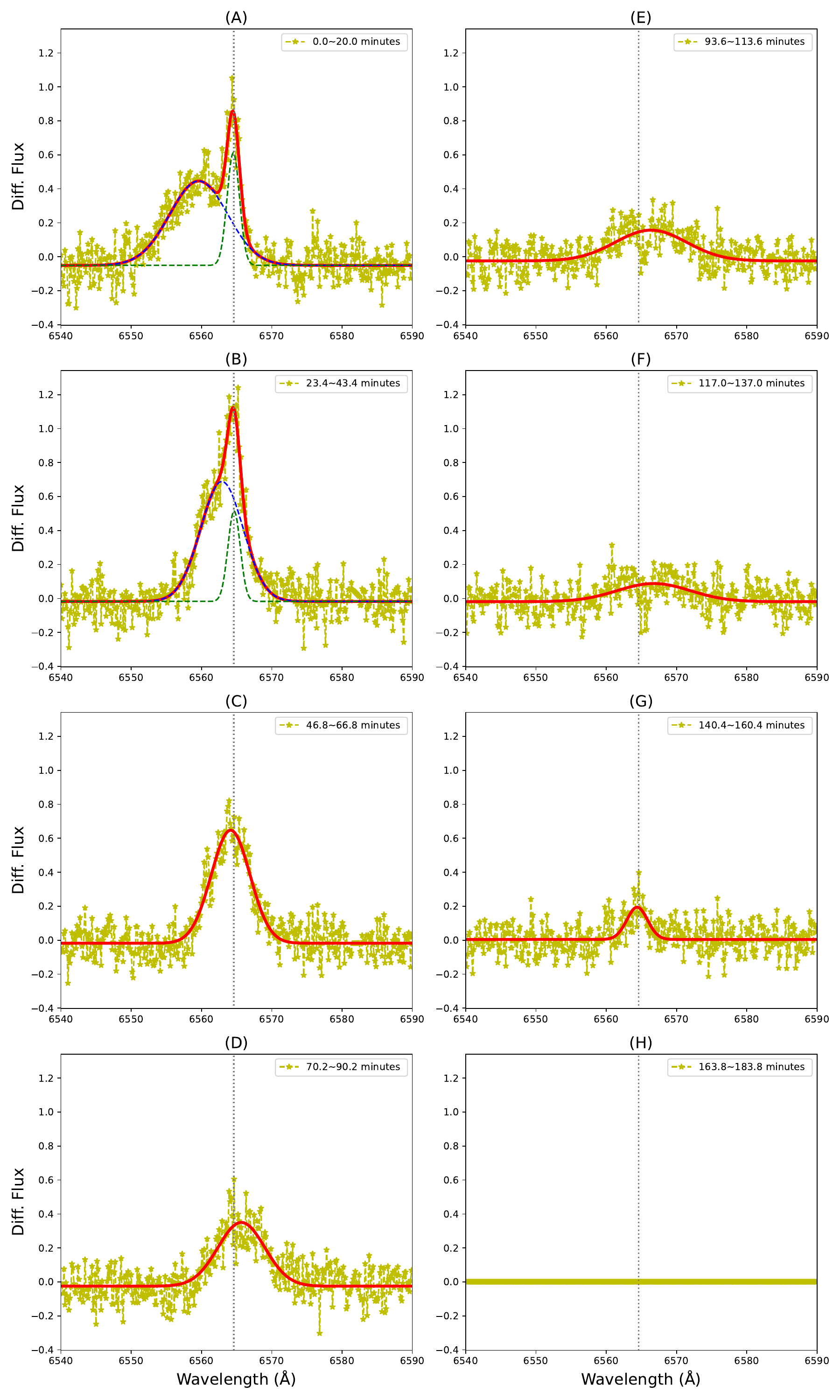}
\caption{Gaussian fitting for the H$\alpha$ line profiles. The yellow star-dashed lines represent the normalized H$\alpha$ line profiles with the reference spectrum subtracted, and the red solid lines show the results of single or double Gaussian fitting. The blue and green dashed lines represent the two Gaussian components. The vertical gray dotted line in each panel marks the rest wavelength of the H$\alpha$ line.
\label{fig:Gaussianfit}}
\end{figure}

To analyze the asymmetry of H$\alpha$ line profiles, we performed Gaussian fitting for the normalized H$\alpha$ line profiles with the reference spectrum subtracted during the flare, including both single-Gaussian and double-Gaussian fittings (Appendix \ref{sec:A}). The Gaussian fitting results for the H$\alpha$ line profiles are shown in Figure \ref{fig:Gaussianfit}. Figures \ref{fig:Gaussianfit}(A) and (B) indicate that during the impulsive phase and peak of the flare, the blue wing of the H$\alpha$ line profiles shows significant enhancement. Figure \ref{fig:Gaussianfit}(C) shows a weak blueshift in the single Gaussian fitted line profile. During the flare decay phase, Figures \ref{fig:Gaussianfit}(D), (E), and (F) exhibit a weak redshift in the single Gaussian profile. Figure \ref{fig:Gaussianfit}(G) shows no significant red or blue shifts, with only an enhancement near the core of the H$\alpha$ line. Therefore, this stellar eruption event reveals two distinct plasma motions: Doppler blueshifts in the H$\alpha$ line profiles during the impulsive phase and near the peak, and Doppler redshifts in the H$\alpha$ line profiles during the flare decay phase.

In this stellar eruption event, the Doppler blueshifts were detected in the H$\alpha$ line, appearing during the impulsive phase and near the peak of the flare. Additionally, from the double Gaussian fitting in Figure \ref{fig:Gaussianfit}(A), the bulk velocity of the blue-shifted Gaussian component reaches $-228\pm11$~km~s$^{-1}$, with a maximum blueshift velocity (see Appendix \ref{sec:A}) of $-605\pm15$~km~s$^{-1}$. Based on observations and theoretical models of solar eruptions, this Doppler blueshift is most likely caused by stellar prominence eruptions. The blueshift is unlikely to be caused by chromospheric evaporation during the flare. In solar flares, chromospheric evaporation includes explosive and gentle evaporation. Explosive evaporation typically occurs in the impulsive phase of large flares, where blueshifts may only appear in spectral lines forming at temperatures above $\sim$1 MK (e.g., Fe XII-XXIV ions) \citep[e.g.,][]{2009ApJ...699..968M, 2015ApJ...811..139T, 2015ApJ...811....7L}. During this time, chromospheric lines exhibit redshifts corresponding to chromospheric condensation, with velocities generally no more than tens of km s$^{-1}$. Gentle evaporation usually occurs in the impulsive phase of small flares or the later stages of large flares, with both chromospheric and coronal lines showing small blueshifts of no more than one hundred km s$^{-1}$ \citep[e.g.,][]{2006ApJ...642L.169M, 2019ApJ...879...30L}. Additionally, the Doppler blueshift is unlikely to be caused by magnetic reconnection outflows during the flare. Solar observations show that the spatial scale of reconnection outflows during solar eruptions is much smaller than that of prominence eruptions \citep[e.g.,][]{2003ApJ...594.1068K, 2018ApJ...853L..15L}, and such signals are almost undetectable in full-disk integrated spectra. Therefore, these mechanisms are not considered the primary sources of the observed Doppler blueshifts in this stellar eruption event. Additionally, the rotation velocity of the host star ($v \sin i = 7.1 \, \mathrm{km \, s^{-1}}$) is significantly lower than the velocity of the blueshifted component. The H$\alpha$ blue-wing enhancement cannot be explained by the rotationally modulated emission from a co-rotating prominence \citep{2021PASJ...73...44M}.

The first two H$\alpha$ spectra obtained during this eruption event both exhibit a double-Gaussian profile. The discussion above suggests that the blueshifted component is most likely caused by an erupting prominence associated with the flare. The equivalent width of the blueshifted component increases from 5.1 \AA\ to 6 \AA, potentially due to a heating of the erupting prominence itself. In these two H$\alpha$ line profiles, the Gaussian component near the rest wavelength of the H$\alpha$ line (central component) shows no significant red or blue shift, likely indicating enhanced line emission from the flaring region due to a heating. Additionally, the equivalent width of the central component decreases from 1.5 \AA\ to 1.2 \AA, and the ratio of the equivalent widths of the blueshifted to central components increases from 3.5 to 4.6. This could result either from a slight weakening of flare heating from the rise phase of the flare to its peak, or from the expanding prominence partially obscuring the H$\alpha$ emission from the flare region.

For the Doppler redshifts observed in the H$\alpha$ line profiles during the flare decay phase, Gaussian fitting shows that the bulk velocities of the red-shifted Gaussian components are all below 100 km s$^{-1}$. This redshift phenomenon is likely due to coronal rain during the flare. Solar observations indicate that coronal rain typically occurs during the flare decay phase and that the typical velocity ranges from 30 to 150 km s$^{-1}$ \citep{2014SoPh..289.4117A, 2016ApJ...818..128O, 2020PPCF...62a4016A, 2021RAA....21..255L, 2022A&A...659A.107C}.

\begin{figure}[ht!]
\plotone{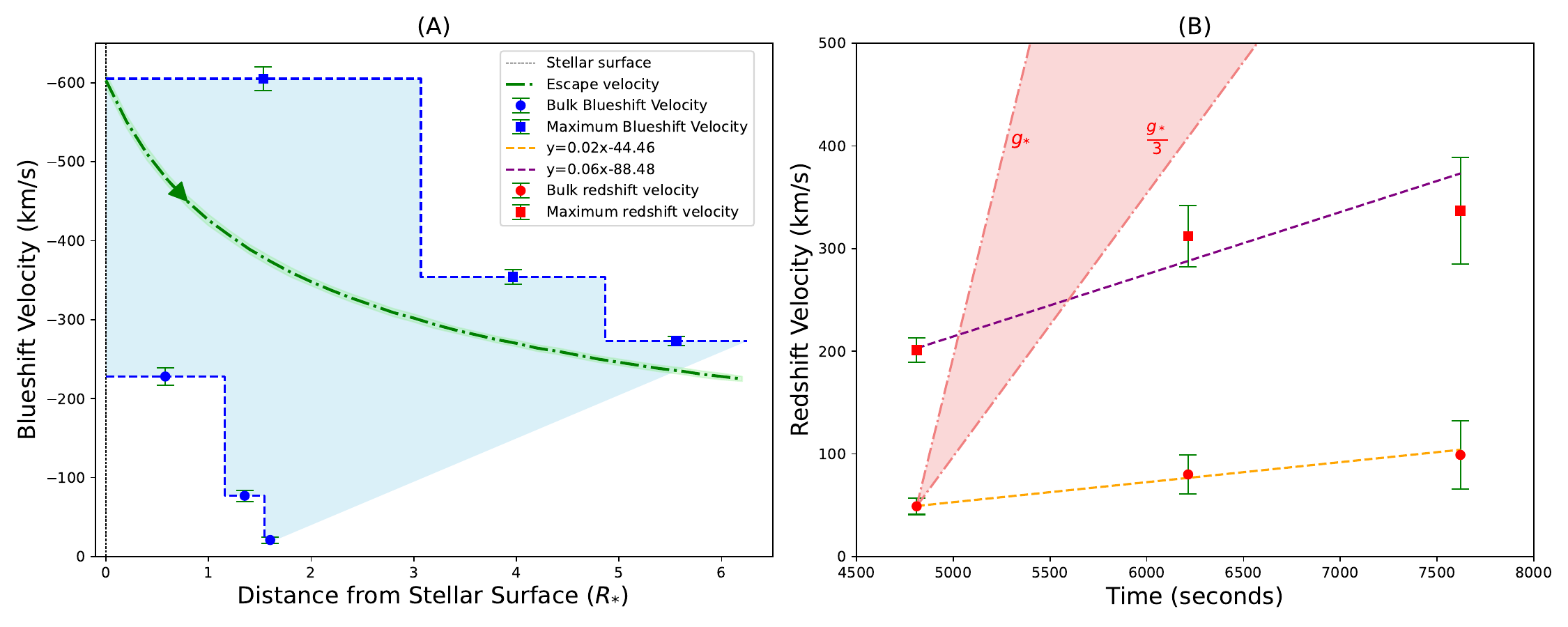}
\caption{Doppler velocity evolution of the blue- and red-shifted Gaussian components in the H$\alpha$ line profiles. Panel (A) shows the ejection distances of the blue-shifted components corresponding to the bulk blue-shift (blue solid circles) and maximum blue-shift velocities (blue squares) within each 20-minute spectral exposure. The sky-blue shaded area represents the possible ejection distances of the prominence plasma, where the ejection velocities range between the bulk blue-shift and maximum blue-shift velocities. The green dashed line indicates the escape velocity as a function of height above the stellar surface. Panel (B) shows the temporal evolution of the bulk red-shift (red solid circles) and maximum red-shift velocities (red squares) of the three red-shifted line profiles. The orange and purple dashed lines represent the linear fitting results for the bulk red-shift and maximum red-shift velocities, respectively. The lightcoral shaded area indicates the range of velocity changes over time, assuming that the initial velocity is the bulk velocity of the first red-shifted line profile and that the acceleration is between one-third and one times the stellar surface gravity.
\label{fig:Vevolution}}
\end{figure}

Figures \ref{fig:Vevolution}(A) and (B) show the evolution of Doppler velocities for the blue- and red-shifted components, respectively. In this stellar eruption event, the blueshift in the H$\alpha$ line persist for up to one hour, suggesting that the blueshift of the first three H$\alpha$ line profiles are related to changes in velocity as the ejected prominence plasma propagates to different heights above the stellar surface. Figure \ref{fig:Vevolution}(A) shows the ejection distances of the blue-shifted components corresponding to the bulk blueshift and maximum blueshift velocities within each 20-minute spectral exposure time. The skyblue shaded area in Figure \ref{fig:Vevolution}(A) represents the possible ejection distances of the prominence plasma within the 20-minute exposure time, where the ejection velocities are between the bulk blueshift and maximum blueshift velocities. Additionally, the green dashed line indicates the escape velocity as a function of height above the stellar surface. Figure \ref{fig:Vevolution}(A) shows that the projected ejection velocity of some prominence plasma along the line of sight exceeds the corresponding escape velocity, suggesting that this prominence eruption may be associated with a stellar CME.

Figure \ref{fig:Vevolution}(B) shows the temporal evolution of the bulk redshift and maximum redshift velocities of the three red-shifted line profiles. The orange and purple dashed lines in the figure represent the linear fitting results of the bulk redshift and maximum redshift velocities, respectively, showing a gradually increasing trend. However, their accelerations are much lower than the typical acceleration of solar coronal rain. The typical acceleration of solar coronal rain is one-third of the solar surface gravitational acceleration \citep{2020PPCF...62a4016A, 2021RAA....21..255L, 2022A&A...659A.107C}. The lightcoral shaded area in the figure indicates the range of velocity changes over time when the initial velocity is assumed to be the bulk velocity of the first red-shifted profile and the acceleration is within the range of one-third to one times the stellar surface acceleration. The much slower increase in redshift velocities of the three H$\alpha$ line profiles may be related to the gradual elevation of post-flare loops, where the coronal rain formed from cooling condensation at the loop tops gains additional acceleration time as it descends along the loops.

\begin{figure}[ht!]
\plotone{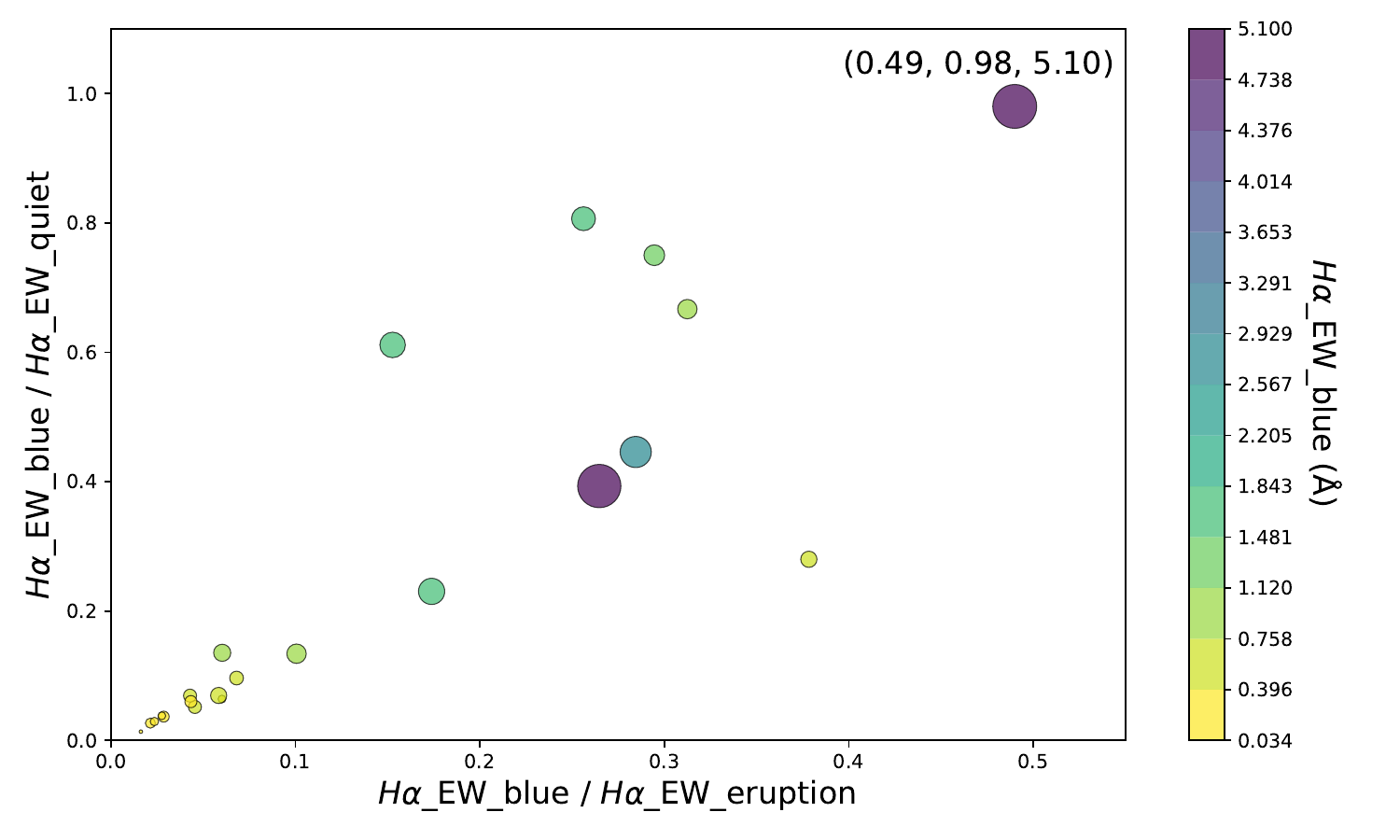}
\caption{Statistical analysis of blue-wing asymmetry in the H$\alpha$ line during stellar flares. The x-axis represents the ratio of the EW of the blue-wing component of the most prominent asymmetric H$\alpha$ line profile to the EW of the H$\alpha$ line profile at the same time (H$\alpha$\_EW\_blue/H$\alpha$\_EW\_eruption). The y-axis represents the ratio of the EW of the blue-wing component of the most prominent asymmetric H$\alpha$ line profile to the EW of the H$\alpha$ line profile during the quiescent period (H$\alpha$\_EW\_blue/H$\alpha$\_EW\_quiet). The size and color of the solid circles represent the EW of the blue-wing component of the most prominent asymmetric H$\alpha$ line profile (H$\alpha$\_EW\_blue). The three values in parentheses indicate the corresponding parameters for the event analyzed in this study.
\label{fig:maxha}}
\end{figure}

\section{Statistical Analysis of Blue-wing Asymmetry in \texorpdfstring{H$\alpha$}{Halpha} Line During Stellar Flares} \label{sec:sec4}

Our observations reveal a significant enhancement of the H$\alpha$ blue wing during the prominence eruption. To compare this event with other stellar flare eruptions, we conducted a statistical analysis of flare-associated prominence eruptions detected through the asymmetry in the blue wing of H$\alpha$. Our sample includes 22 events exhibiting clear H$\alpha$ blue wing asymmetries: one event associated with H$\alpha$ blue wing absorption possibly related to a stellar filament eruption \citep{2022NatAs...6..241N} and 21 events associated with H$\alpha$ blue wing enhancement due to possible prominence eruptions \citep[and this work]{2016A&A...590A..11V, 2018PASJ...70...62H, 2019A&A...623A..49V, 2020MNRAS.499.5047M, 2021PASJ...73...44M, 2021A&A...646A..34K, 2022A&A...663A.140L, 2023ApJ...948....9I, 2024PASJ...76..175I, 2024ApJ...961...23N}.

For each event in this sample, we obtained three parameters: (1) the EW of the blue-wing component of the most prominent asymmetric H$\alpha$ line profile (H$\alpha$\_EW\_blue), (2) the ratio of the EW of the blue-wing component to the EW of the H$\alpha$ line profile at the same time (H$\alpha$\_EW\_blue/H$\alpha$\_EW\_eruption), and (3) the ratio of the EW of the blue-wing component to the EW of the H$\alpha$ line profile during the quiescent period (H$\alpha$\_EW\_blue/H$\alpha$\_EW\_quiet). Figure \ref{fig:maxha} shows that all three parameters for the prominence eruption in this study are the highest in the sample: (1) the EW of the H$\alpha$ blue wing component in this stellar prominence eruption is the largest observed to date; (2) the H$\alpha$\_EW\_blue/H$\alpha$\_EW\_eruption ratio is the highest recorded; and (3) the EW of the H$\alpha$ blue wing component is comparable to the EW of the H$\alpha$ line profile during the quiescent period of the host star. These results suggest that we observed an exceptionally intense stellar prominence eruption, possibly accompanied by a CME capable of causing significant disturbances to the space environment surrounding the host star.

\section{Analysis of Asymmetric \texorpdfstring{H$\alpha$}{Halpha} Line Profile Based on the Two-Cloud Model}\label{sec:sec5}

\begin{figure}[ht!]
\plotone{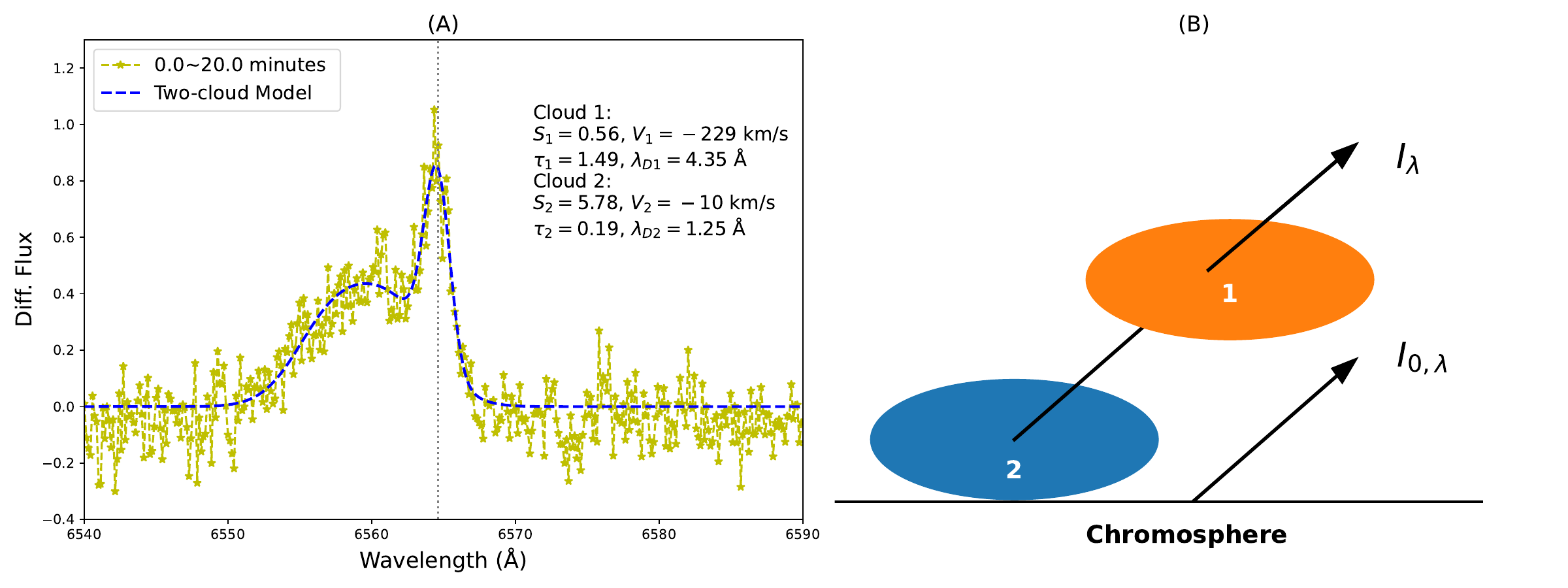}
\caption{Fitting the most asymmetric H$\alpha$ line profile using the two-cloud model. Panel (A) shows the fitting result, where the yellow star-dashed line represents the normalized H$\alpha$ line profile with the reference spectrum subtracted, and the blue dashed line shows the two-cloud model fit. The fitting parameters are indicated in the upper right corner, with Cloud 1 representing the erupting prominence and Cloud 2 representing the flare region. Panel (B) shows a schematic diagram of the two-cloud model.
\label{fig:cloudmodel}}
\end{figure}

The multi-cloud model is commonly used to analyze asymmetric spectral line profiles during solar flares \citep[e.g.,][]{2001A&A...380..704G, 2014ApJ...792...13H}. In our observation, the H$\alpha$ line profile during the flare impulsive phase exhibits a significant blue-wing enhancement. To analyze the physical parameters of the erupting prominence, we propose a two-cloud model (Appendix \ref{sec:C}) combined with the least squares method to fit the H$\alpha$ line profile with the most prominent blue-wing enhancement (the first profile). The fitting results are shown in Figure \ref{fig:cloudmodel}(A), with a reduced chi-square ($\chi^2_\nu$) \citep{2001A&A...374.1108P, 2011ApJ...738...18T} of 1.66, indicating a overall good fit. A corresponding schematic of the cloud model is shown in Figure \ref{fig:cloudmodel}(B).

The two-cloud model fitting result shows that the source function $S_1$ of the erupting prominence (cloud 1) is 0.56, with an optical depth $\tau_1$ of 1.49. The Doppler shift along the line of sight $V_1$ is -229 km s$^{-1}$ (negative indicating blueshift), and the Doppler broadening $\Delta \lambda_{D,1}$ is 4.35 \AA. For the flare region (cloud 2), the source function $S_2$ is 5.78, with an optical depth $\tau_2$ of 0.19, a Doppler shift $V_2$ of -10 km s$^{-1}$, and a Doppler broadening $\Delta \lambda_{D,2}$ of 1.25 \AA. Based on the fitting result, the excitation temperature $T_{\text{ex1}}$ of the erupting prominence (cloud 1) is 5482 K, and the column number density of hydrogen atoms at the second level along the line of sight $N_{2,1}$ is $4.71 \times 10^{13}$\,cm$^{-2}$. For the flare region (cloud 2), the excitation temperature $T_{\text{ex2}}$ is 13087 K, and the column number density $N_{2,2}$ is $1.75 \times 10^{12}$\,cm$^{-2}$. The excitation temperatures of the erupting prominence (cloud 1) and flare region (cloud 2) are 167\% and 399\% of the stellar surface effective temperature (3283 K), respectively. Additionally, based on the parameters of the host star and the result of the two-cloud model, the projected area of the erupting prominence along the line of sight during the 20-minute exposure of the first spectrum is estimated to be $1.19 \times 10^{18} \, \text{m}^2$, approximately 6.8 times the stellar disk area ($A_{\text{star}} = \pi R_{\text{star}}^2$). Based on the cloud model result, the mass of the erupting prominence is inferred to range from $1.6 \times 10^{19} \, \text{g}$ to $7.2 \times 10^{19} \, \text{g}$ (Appendix \ref{sec:D}). Compared to solar eruptive filaments, which have masses on the order of $10^{12} \, \text{g}$ to $10^{16} \, \text{g}$ \citep[e.g.,][]{2010SSRv..151..243L, 2011LRSP....8....1C}, this stellar prominence eruption is a monster prominence eruption. More importantly, the mass ratio of the erupting prominence to its host star is the largest among all reported stellar prominence eruptions/CMEs \citep[e.g.,][]{1990AA...238..249H, 2019A&A...623A..49V, 2021PASJ...73...44M, 2022NatAs...6..241N}.

\section{Summary} \label{sec:sec6}

In this study, we report the detection of an extreme stellar prominence eruption on an M dwarf LAMOST J044431.62+235627.9 based on LAMOST H$\alpha$ time-domain spectral observations. This eruption was accompanied by a superflare with H$\alpha$ band energy exceeding $4.6 \times 10^{31}$ erg and a duration of over 160.4 minutes. The H$\alpha$ time-domain spectra of the flare show significant blue-wing enhancements during the impulsive phase and near the peak, as well as redshifted features during the decay phase of the flare. The projected bulk blueshift velocity and maximum blueshift velocity reached $-228\pm11$~km~s$^{-1}$ and $-605\pm15$~km~s$^{-1}$, respectively. Our analysis indicates that the blueshift is most likely associated with a stellar prominence eruption during the flare, while the redshift is caused by coronal rain.

Importantly, our analysis of the prominence velocities at various heights above the surface of the host star indicates that at least some materials within the prominence should exceed the corresponding escape velocities, suggesting a potential stellar CME. The significantly enhanced blue wing of the H$\alpha$ line indicates this is an extreme prominence eruption. The EW of the excess blue wing component in this prominence eruption is the largest observed to date. The proportion of the EW of the most significant blue wing component to the EW of the H$\alpha$ line profile at the same time is the highest recorded. The EW of the asymmetric H$\alpha$ blue wing component is comparable to that of the H$\alpha$ line profile during the quiescent period of the host star.

By applying a two-cloud model fit to the H$\alpha$ line profile with the most significant blue-wing asymmetry, we obtained an excitation temperature of 5482 K for the eruptive prominence (cloud 1), and 13087 K for the flare region (cloud 2). Additionally, during the 20-minute exposure time of this line profile, the projected area of the eruptive prominence along the line of sight was found to be approximately 6.8 times the stellar disk area ($A_{\text{star}} = \pi R_{\text{star}}^2$), and the estimated mass of the eruptive prominence ranged from $1.6 \times 10^{19}$ g to $7.2 \times 10^{19}$ g. The mass ratio of this erupting prominence to its host star is the largest among all reported stellar prominence eruptions/CMEs. With increasing awareness that stellar CMEs or prominence eruptions may play a critical role in shaping the habitability of exoplanetary environments, the detection of such an extraordinary prominence eruption should provide important observational constraints for research of exoplanet habitability and astrobiology.

\begin{acknowledgments}
We sincerely thank the referee for providing constructive feedback and valuable insights. This work is supported by the National Natural Science Foundation of China (NSFC) under grants 12250006, 12425301, 12473055, 12103004, and by the Guizhou University Natural Science Special Research Fund (Special Post) under project number 202358. L.Y.Z. is supported by the NSFC grant 12373032. H.C.C. is supported by the NSFC grant 12103005. Y.L. is supported by the Strategic Priority Research Program of the Chinese Academy of Sciences, Grant No. XDB0560000. The Guoshoujing Telescope (the Large Sky Area Multi-Object Fiber Spectroscopic Telescope, LAMOST) is a National Major Scientific Project built by the Chinese Academy of Sciences. Funding for the LAMOST project has been provided by the National Development and Reform Commission, and it is operated and managed by the National Astronomical Observatories, Chinese Academy of Sciences.
\end{acknowledgments}

%



\appendix
\renewcommand{\thefigure}{A\arabic{figure}} 
\setcounter{figure}{0} 

\section{Stellar Superflares Observed by TESS}\label{sec:1A}
\begin{figure}[ht!]
\plotone{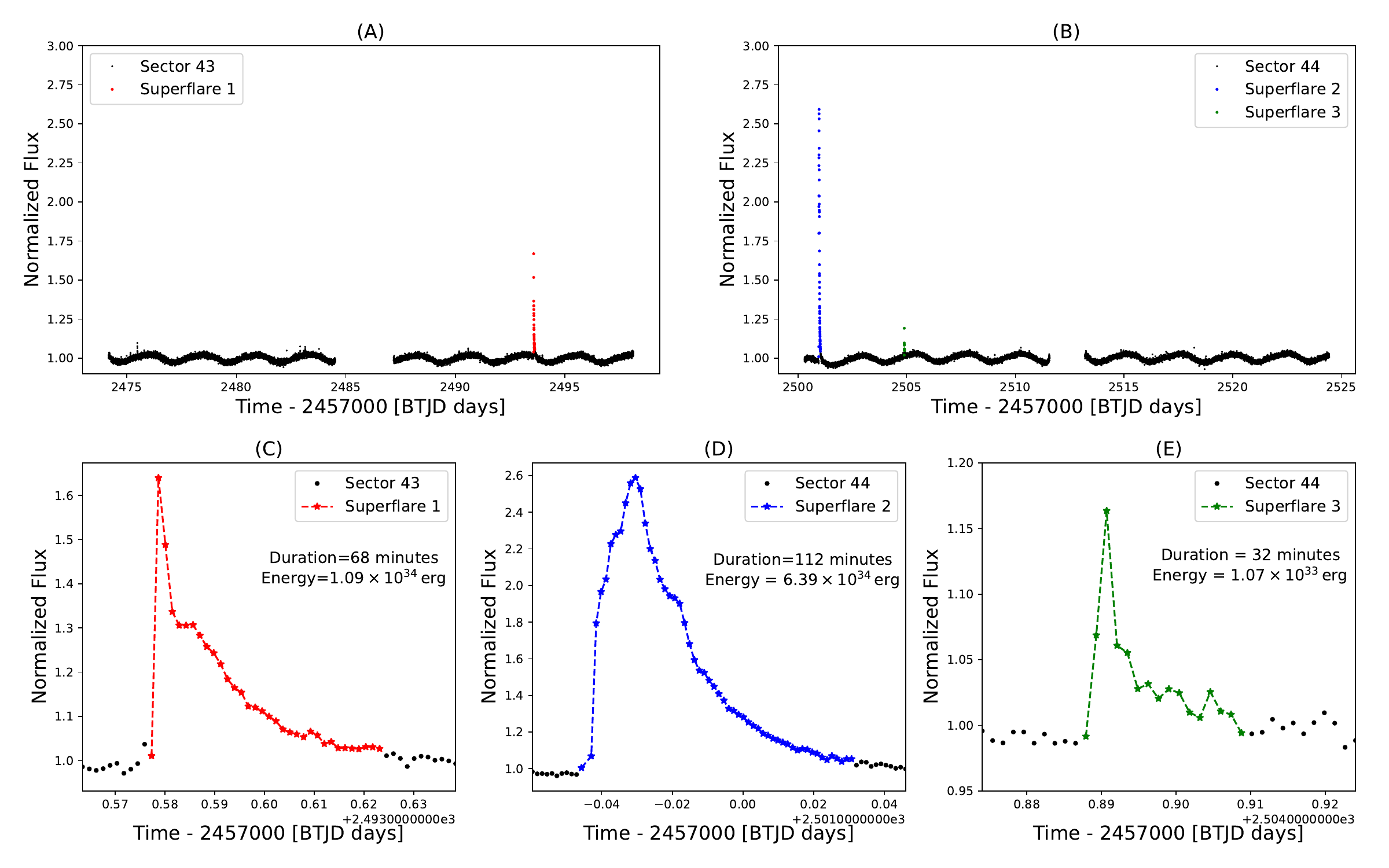}
\caption{Stellar superflares of LAMOST J044431.62+235627.9 observed by TESS. (A) and (B): Two sets of TESS photometric data with three superflares marked by red, blue and green stars. (C), (D) and (E): TESS light curve profiles of the three superflares.
\label{fig:tess_flares}}
\end{figure}

To verify the activity of LAMOST J044431.62+235627.9, we retrieved two sets of photometric data for this M dwarf from the TESS database. The flux-normalized TESS photometric data are shown in panels (A) and (B) of Figure \ref{fig:tess_flares}. After analyzing both datasets with AltaiPony \citep{2021A&A...645A..42I}, we detected three stellar superflares. The calculation of the bolometric energy of these superflares follows the method of \cite{2013ApJS..209....5S}. The light curves of the three superflares are displayed in panels (C), (D), and (E) of Figure \ref{fig:tess_flares}, where the flare duration and energy are indicated. Based on these TESS observations, we calculated the superflare occurrence rate of this M dwarf to be one event every 14.6 days. Additionally, a Lomb-Scargle analysis of the TESS data reveals a rotation period of 2.43 days for this M dwarf.

\section{Gaussian fitting of \texorpdfstring{$H\alpha$}{Halpha} line profiles}\label{sec:A}

To obtain information about the Doppler shifts of the H$\alpha$ line, Gaussian fitting was performed on the normalized H$\alpha$ profiles with the reference spectrum subtracted (Diff. Flux) during the flare. This fitting included both single-Gaussian and double-Gaussian fittings, as described by the following equations.

For single Gaussian fitting:
\begin{equation}
\text{Diff. Flux} = A_0 + A_1 \exp \left( -\frac{(\lambda - \lambda_1)^2}{2 \sigma_1^2} \right)
\end{equation}

For double Gaussian fitting:
\begin{equation}
\text{Diff. Flux} = A_0 + A_1 \exp \left( -\frac{(\lambda - \lambda_1)^2}{2 \sigma_1^2} \right) + A_2 \exp \left( -\frac{(\lambda - \lambda_2)^2}{2 \sigma_2^2} \right)
\end{equation}

The bulk velocity and maximum velocity of the Gaussian components were calculated using:
\begin{equation}
V_{i,\text{bulk}} = \frac{(\lambda_i - \lambda_0) \cdot c}{\lambda_0}, \quad i = 1, 2
\end{equation}
\begin{equation}
V_{i,\text{max}} = \frac{(\lambda_i - \lambda_0 \pm 2\sigma_i) \cdot c}{\lambda_0}, \quad i = 1, 2
\end{equation}
Here, $\lambda_{0}$ is the rest wavelength of the H$\alpha$ line in air (6564.6 \AA), and $c$ is the speed of light. When analyzing the Doppler redshift of the H$\alpha$ line, the maximum redshift velocity is determined at a distance of one $\sigma$ from the center of the redshifted component, due to the decreasing signal-to-noise ratio of the redshifted component.

\section{Two-Cloud Model}\label{sec:C}

The following equations describe the two-cloud model used for fitting, where the upper cloud (cloud 1) represents the erupting prominence and the lower cloud (cloud 2) represents the flare region:
\begin{equation}
I_\lambda = S_1 \cdot \left(1 - \exp{(-\tau_{\lambda,1})}\right) + S_2 \cdot \left(1 - \exp{(-\tau_{\lambda,2})}\right) \cdot \exp{(-\tau_{\lambda,1})}
\end{equation}
where $I_{\lambda}$ is the observed intensity, and $S_1$, $S_2$, $\tau_{\lambda,1}$, and $\tau_{\lambda,2}$ are the source functions and optical depths of the clouds, respectively. The optical depth $\tau_{\lambda,j}$ as a function of wavelength $\lambda$ is given by:
\begin{equation}
\tau_{\lambda,j} = \tau_j \cdot \exp \left( -\left( \frac{\lambda - \lambda_0 - \lambda_0 V_j / c}{\Delta \lambda_{D,j}} \right)^2 \right), \quad j=1,2
\end{equation}
where $\tau_{j}$ is the optical depth at the central wavelength for the $j$-th cloud, $V_{j}$ and $\Delta \lambda_{D,j}$ are the Doppler shift velocity and Doppler broadening along the line of sight, $\lambda_{0}$ is the rest wavelength of the H$\alpha$ line in air (6564.6 \AA), and $c$ is the speed of light. For the H$\alpha$ line profiles observed during stellar flares, we first normalized the continuum and subtracted the reference spectrum (quiescent spectrum). Therefore, the background radiation of the chromosphere was not considered in the two-cloud model. A schematic of the cloud model is shown in Figure \ref{fig:cloudmodel}(B).

Additionally, the excitation temperature $T_{\text{ex}}$ and the column number density of hydrogen atoms at the second level along the line of sight $N_{2}$ for each cloud can be calculated using the following relationships \citep{1995ASSL..199.....T, 2001A&A...380..704G}:
\begin{equation}
T_{\text{ex}} = \frac{hc}{\lambda_0 k} \left/ \ln{\left( 1 + \frac{2hc^2}{\lambda_0^5 S} \cdot \frac{b_3}{b_2} \right)} \right.
\end{equation}
\begin{equation}
N_2 = \frac{m_e c^2 \tau_0 \Delta \lambda_D}{\sqrt{\pi} e^2 \lambda_0^2 f_{2,3}}
\end{equation}
where $h$ is the Planck constant, $k$ is the Boltzmann constant, $b_{2}$ and $b_{3}$ are departure coefficients from local thermodynamic equilibrium (LTE), and $f_{2,3}$ is the oscillator strength of the H$\alpha$ line. $m_{e}$ and $e$ are the electron mass and charge, respectively. Excitation temperature is the temperature that describes the distribution of particles among energy levels based on a Boltzmann distribution.

\section{Estimation of Prominence Parameters}\label{sec:D}

Using the stellar parameters mentioned in Section \ref{sec:sec2}, the line-of-sight H$\alpha$ luminosity of the erupting prominence $L_{\text{H}\alpha\_\text{bluewing\_los}}$ can be calculated using the following equation:
\begin{equation}
L_{\text{H}\alpha\_\text{bluewing\_los}} = \chi_{\text{H}\alpha} \cdot L_{\text{bol}} \cdot \frac{\text{EW}_{\text{H}\alpha\_\text{bluewing}}}{4}
\end{equation}
where $\chi_{\text{H}\alpha}$ is the ratio between the continuum flux near the H$\alpha$ line and the bolometric flux \citep{2004AJ....128..426W, 2018MNRAS.476..908F}. $\text{EW}_{\text{H}\alpha\_\text{bluewing}}$ is the EW of the H$\alpha$ blue-wing component. In this event, the EW of the most prominent blue-wing component $\text{EW}_{\text{H}\alpha\_\text{bluewing}}$ is 5.1 \AA. The calculated $L_{\text{H}\alpha\_\text{bluewing\_los}}$ is $1.99 \times 10^{27}$\,erg s$^{-1}$.

From the two-cloud model fitting, the source function $S_{1}$ of the erupting prominence (cloud 1) is 0.56, leading to the radiance $I_{\text{H}\alpha\_\text{los}}$ of the prominence along the line of sight being $1.67024 \times 10^5 \, \text{erg} \, \text{s}^{-1} \, \text{cm}^{-2}$. Dividing the $L_{\text{H}\alpha\_\text{bluewing\_los}}$ by $I_{\text{H}\alpha\_\text{los}}$, we estimated the projected area $A_{\text{CME\_los}}$ of the erupting prominence along the line of sight during the 20-minute exposure of the first spectrum to be $1.19 \times 10^{18} \, \text{m}^2$, which is approximately 6.8 times the stellar disk area ($A_{\text{star}} = \pi R_{\text{star}}^2$).

Based on solar observations and theoretical models, there is an approximate relationship between the electron density $n_{e}$ and the $N_{2}$ for the prominence \citep{1971SoPh...18..391P, 2001A&A...380..704G}:
\begin{equation}
N_2 = n_2 \cdot D,
\end{equation}
\begin{equation}
n_e \approx 3.2 \times 10^8 \cdot \sqrt{n_2}
\end{equation}
where $n_{2}$ is the density of hydrogen atoms in a unit volume of the second level along the line of sight, and $D$ is the average thickness of the erupting prominence along the line of sight. Assuming that the prominence expands into a spherical volume, we calculated the radius $R_{\text{CME}}$ of the prominence from the projected area $A_{\text{CME\_los}}$ to be $2.6 \cdot R_{\text{star}}$, which is taken as $D$. Our calculation yielded that the electron density $n_{e}$ is approximately $9 \times 10^9$ cm$^{-3}$. Additionally, for solar prominences, the hydrogen density $n_{H}$ and electron density $n_{e}$ follow the relationship \citep{2010SSRv..151..243L, 2023ApJ...948....9I}:
\begin{equation}
\frac{n_e}{n_H} \sim 0.2 - 0.9.
\end{equation}
Therefore, the mass of the erupting prominence $M_{\text{CME}}$ can be estimated as:
\begin{equation}
M_{\text{CME}} \sim m_H n_H \frac{4}{3} \pi R_{\text{CME}}^3,
\end{equation}
ranging from $1.6 \times 10^{19}$\,g to $7.2 \times 10^{19}$\,g. The kinetic energy of the eruptive prominence is estimated to range from \(4.2 \times 10^{33}\ \text{erg}\) to \(1.9 \times 10^{34}\ \text{erg}\).


\bibliography{References}{}
\bibliographystyle{aasjournal}



\end{document}